\documentclass[aip,apl,reprint, superscriptaddress]{revtex4-1}

\usepackage{graphicx}
\usepackage[ansinew]{inputenc}
\usepackage{array}
\usepackage{color}
\usepackage{amsmath}
\usepackage{amsxtra}
\usepackage{amstext}
\usepackage{amssymb}
\usepackage{latexsym}
\begin{document}

\title{Mapping the Dirac point in gated bilayer graphene}

\author{A.	Deshpande}
\author{B. J. LeRoy}
\email{leroy@physics.arizona.edu}
\affiliation{Department of Physics, University of Arizona, Tucson, AZ, 85721 USA.}

\author{W. Bao}
\author{Z. Zhao}
\author{C. N. Lau}
\affiliation{Department of Physics, University of California at Riverside, Riverside, CA 92521 USA.}

\date{\today}

\begin{abstract}
We have performed low temperature scanning tunneling spectroscopy measurements on exfoliated bilayer graphene on SiO$_2$. By varying the back gate voltage we observed a linear shift of the Dirac point and an opening of a band gap due to the perpendicular electric field.  In addition to observing a shift in the Dirac point, we also measured its spatial dependence using spatially resolved scanning tunneling spectroscopy. The spatial variation of the Dirac point was not correlated with topographic features and therefore we attribute its shift to random charged impurities.
\end{abstract}

\maketitle
Monolayer graphene (MLG), which is just a single sheet of carbon atoms thick, has novel electronic properties as a consequence of its linear band structure. Stacking one more layer on top of the monolayer gives rise to bilayer graphene (BLG) that is an exciting system with a different set of tunable properties \cite{novoselov2007,neto2009}. The bilayer structure is characterized by a quadratic dispersion relation $E=\pm\hbar^2k^2/2m$ with the conduction band and valence bands touching making BLG a zero band gap semiconductor. When an electric field is applied perpendicular to the plane of carbon atoms, it is possible to open up a band gap between the conduction band and valence band \cite{mccann2006,castro2007,min2007}. Recent experiments with techniques like angle resolved photoemission spectroscopy \cite{ohta2006}, infrared spectroscopy \cite{li2008,li2009,zhang2009} and  transport measurements with a double-gate \cite{oostinga2008} have confirmed this band gap opening. These techniques are non-local and only provide information about the average properties of the BLG. However, from a device application perspective it is important to get details about how the spatial extent and morphology of the layers affect the electronic properties. Scanning tunneling microscopy (STM) is a powerful tool for this purpose. Previous STM studies have shown that impurites in MLG \cite{deshpande2009,zhangAOP2009}and phonons \cite{zhang2008}influence the charge-carrier scattering mechanisms in graphene. In this letter, we present scanning tunneling spectroscopy results for BLG on a SiO$_2$ substrate.  These results show the spatial variation of the Dirac point as well as the control of the Dirac point and band gap due to the application of an electric field from the back gate.

The BLG was prepared using the mechanical exfoliation technique\cite{novoselov2004,novoselov2005}. Degenerately doped Si with 300 nm thick SiO$_2$ on top was used as a back gate. Bilayer areas were identified using an optical microscope and then Ti/Au electrodes were deposited using a shadow mask technique described elsewhere \cite{LauPrep}. The device was then cooled to 4.6 K using an Omicron low temperature STM operating in ultrahigh vacuum (p $\leq 10^{-11}$ mbar). Electrochemically etched tungsten tips that exhibited a constant density of states on a Au surface were used for imaging and spectroscopy to avoid unwanted tip effects.  Due to the cleaner fabrication procedure, no PMMA is used, it is possible to obtain atomic resolution images over large areas of the BLG without any additional cleaning procedure unlike in previous STM measurements on exfoliated graphene \cite{deshpande2009,zhang2008,zhangAOP2009,ishigami2007,stolyarova2007,geringer2009}.

\begin{figure}[b]
\includegraphics[width=0.45 \textwidth]{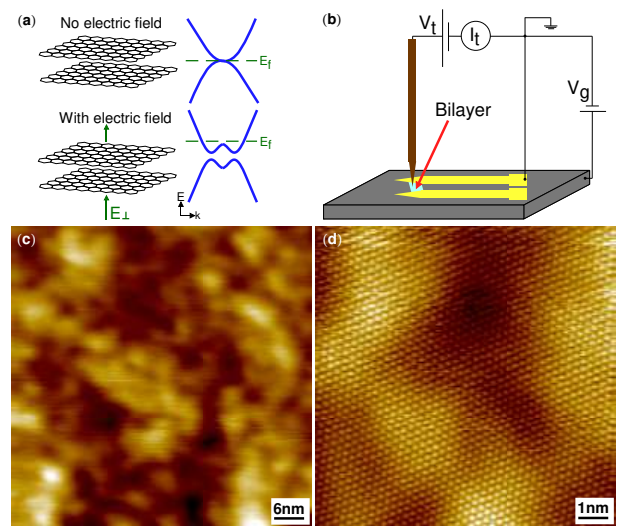}
\caption{\label{fig:schematic} (color online) (a) Illustration of bilayer graphene and its band structure with and without a perpendicular electric field.  The electric field shifts the Dirac point and opens up a gap. (b) Schematic of mechanically exfoliated bilayer graphene flake showing gold electrodes on a SiO$_2$ substrate with STM and back gate connections. (c) Constant current STM image of bilayer graphene (0.5 V, 100 pA, 60 nm $\times$ 60 nm). (d) Atomic resolution showing the triangular lattice (0.5 V, 100 pA, 10 nm $\times$ 10 nm)}
\end{figure}

Figure 1(a) illustrates the concept of band gap opening for BLG. The measurement set up with STM and back gate connections is shown in Fig. 1(b). The BLG is grounded and voltages are applied to the STM tip and the back gate.  As the voltage is increased on the back gate or tip, an electrical field perpendicular to the BLG is established which shifts the Dirac point and opens a band gap.  Figure 1(c) is a 60 nm $\times$ 60 nm STM image of the BLG showing modulations due to the underlying SiO$_2$ substrate as typically seen in MLG samples on a SiO$_2$ substrate \cite{deshpande2009,ishigami2007,stolyarova2007,geringer2009, zhang2008, zhangAOP2009}. The triangular lattice arises due to the presence of two layers that leads to only a single sublattice being observed in the STM images \cite{stolyarova2007}.

\begin{figure}[b]
\includegraphics[width=0.45 \textwidth]{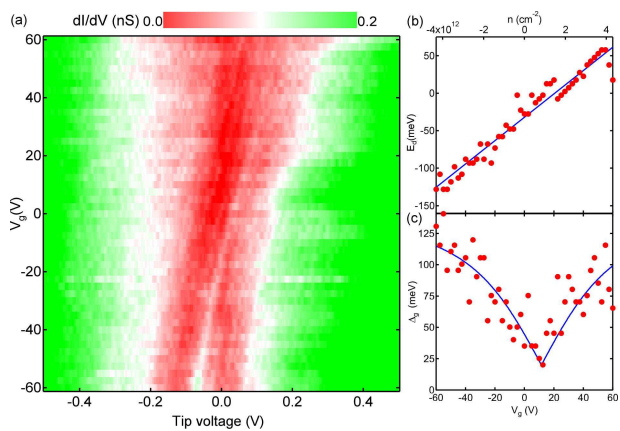}
\caption{\label{fig:2Dpoint} (color online) (a) Point spectroscopy measurements of BLG as a function of back gate and tip voltage (energy). The color scale represents the local density of states, dI/dV, with red areas low and green areas high.  The minimum in the density of states corresponds to the Dirac point and it shifts linearly with gate voltage.  Similarly, the width of the minimum increases with increasing gate voltage indicating that a band gap is opening. (b) Position of the Dirac point E$_d$ determined from the dI/dV curves in (a) as a function of the gate voltage V$_g$ (bottom) and induced carrier density $n$ (top). A straight line fit to the data is shown. (c) Band gap variation with V$_g$. Dotted line is the fit (see text) to the band gap variation. }
\end{figure}

Figure 2(a) shows a plot of dI/dV as a function of tip and gate voltage.  The dI/dV point spectroscopy measurements are performed using standard lockin techniques with an ac modulation of 5 mV rms at 574 Hz.  As we sweep through different gate voltages from -60 V to +60 V in steps of 2.5 V,  two distinct observations can be made.  The location of the minimum in the dI/dV curves shifts towards positive tip voltage with increasing gate voltage.  Also, the width of the red region which corresponds to low values of dI/dV increases with an increasing gate voltage.  In addition, there is an extra peak within the gap (white area visible at negative gate voltages) but further investigation is necessary to attribute it to any of theoretically predicted features like impurities and midgap states \cite{nilsson2007}.

The gate voltage induces electrons or holes into BLG and generates an electric field perpendicular to the plane of the bilayer. It leads to an uneven induced charge on each layer and creates an electrostatic potential difference between the two layers that screens the external field and opens a band gap as predicted by the calculations \cite{mccann2006,castro2007,min2007}. The red region in Fig. 2(a) shows the band gap opening with gate voltage and the shift of the Dirac point. From the dI/dV curves we have determined the band gap at each gate voltage and the energy of the Dirac point based on the position of the minimum in the curve.

A plot of the energy of the Dirac point as a function of the gate voltage is shown in Fig. 2(b). The variation of Dirac point with gate voltage is linear with a slope of 1.56 meV/V.  This linear dependence on gate voltage is different from MLG which has a square root dependence due to the linear bandstructure\cite{zhang2008}.  The linear dependence is expected for BLG because the bandstructure is quadratic $E = \hbar^2 k^2 / 2m^*$ where $m^*$ is the effective mass of electrons in BLG. For a two-dimensional system, the electron density is written as $n=k^2/\pi$.  Since the back gate and the BLG act as a parallel plate capacitor, the induced charge on the BLG is given by $n = \frac {\epsilon \epsilon_0 V_g}{e t_{ox} }$ where $t_{ox}$ is the thickness of the oxide layer. Putting these equations together gives the shift in energy due to the back gate as $E = \hbar^2 \pi \epsilon  \epsilon_0 V_g / 2 {m^*} e t_{ox}$. A second effect of the gate voltage is that it leads to an electric field between the two layers of graphene. Assuming no screening, i.e each layer carries charge density \cite{castro2007} $n/2$, this field creates a potential difference between the two layers, $V = \frac{\epsilon d V_g }{2 e t_{ox} }$ where $d$ is the interlayer separation, 0.335 nm \cite{min2007} and the dielectric constant between the layers is 1 . This potential difference reduces the effective voltage between the tip and the BLG.  The net effect is that the energy of the Dirac point as observed by the tip shifts as
$$E = \frac{\epsilon V_g}{2 e t_{ox} }\left(\frac{\hbar^2 \pi \epsilon_0}{m^*}- d \right)$$

Using the above equation we find that the value of effective mass that fits our observed shift of 1.56 meV/V is $0.023m_e$ where $m_e$ is the mass of the electron.  This value of the effective mass is slightly lower than tight-binding calculations which give $m^*=0.033 m_e$, however many body effects can reduce this mass\cite{kusminskiy2009}.  Furthermore, our analysis did not take into account the electric field due to the tip which would tend to lessen the effect of the gate and therefore increase the effective mass needed to fit our data.  We have also neglected screening between the layers which reduces the potential difference between the layers \cite{mccann2006,castro2007}.  These effects could explain the discrepancy between our value and the calculated values of the effective mass.

Figure 2(c) shows the band gap as a function of gate voltage V$_g$ and carrier density $n$. The band gap has a finite minimum value around a gate voltage of 15 V.  This is the voltage needed to have the Dirac point at the Fermi energy and therefore the minimum charge on the BLG.  The magnitude of the band gap is fit using the equation \cite{castro2007} $\Delta_g=[e^2V^2t_{\bot}^2/(t_{\bot}^2+e^2V^2)]^{1/2}$ where V is the potential between the layers and the only free parameter is $t_{\bot}$ the interplane hopping.  The best fit occurs for a value of $t_{\bot}=0.12$ eV which gives an effective mass of $0.017m_e$.  Once again we have not taken into account the screening between layers which will reduce the potential difference and thus give a larger effective mass for our fit.

\begin{figure}[t]
\includegraphics[width=0.45 \textwidth]{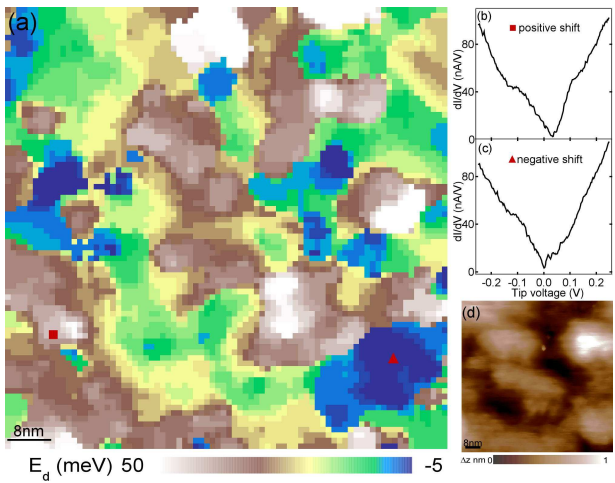}
\caption{\label{fig:map} (color online) (a) A 2D map (80 nm $\times$ 80 nm) showing the spatial variation of the Dirac point calculated from a dI/dV map of bilayer graphene. The colors correspond to positive and negative shift of the Dirac point. (b) dI/dV curve from the area of positive shift shown in (a) by a filled square. (c) dI/dV curve from the area of negative shift shown in (a) by a filled triangle.(d) Topography of the area shown in (a),(0.5 V, 100 pA, 80 nm $\times$ 80 nm), the color scale shows the height variation of 1 nm. }
\end{figure}

In the case of MLG the shift of the Dirac point as a function of position leads to the formation of electron and hole puddles that can be seen in density of states measurements \cite{deshpande2009,zhangAOP2009,martin2008}. To measure the spatial variation of the Dirac point for BLG, we have recorded LDOS maps for large areas of the flake, typically 80 nm $\times$ 80 nm. From the individual spectroscopy curves in the LDOS map we have calculated the energy of the Dirac point.  This gives the energy of the Dirac point as a function of position in the BLG.   Figure 3(a) shows the Dirac point for one such LDOS map. The color scale corresponds to the energy of the Dirac point. A dI/dV curve from an area of positive shift (red square in Fig. 3(a)) is shown in Fig. 3(b). Figure 3(c) shows a dI/dV curve from an area of negative shift (red triangle in Fig. 3(a)). The energy of the Dirac point is determined by the location of the minimum in these curves.  The energy of the Dirac point is not evenly centered around E$_f$ as the average value is at $\overline{E_d}= 20$ meV. Figure 3(d) shows the topograph corresponding to the region in Fig. 3(a). The color scale here corresponds to the height variation of the image. We can clearly see from Fig. 3(a) and 3(d) that there is no correlation between the spatial variation of the Dirac point and the features in the topograph. Hence we claim that the origin of the Dirac point shift, at zero V$_g$ when no charge carriers are induced, lies in the random charged impurities as seen in the case of MLG. The impurity density variation can be calculated from the average measured shift in the Dirac point $\overline{E_d}$ using $\overline{n} = \frac{2 m^* \overline{E_d}}{\pi \hbar^2}$ to be $3.8\times 10^{11}$cm$^{-2}$. These results are consistent with an electrical transport measurement on the BLG which shows the Dirac point shifted slightly away from $V_g=0$.  However, electrical transport measurements give only one number for the Dirac point and do not give any information about its spatial variation.

In conclusion, we have presented scanning tunneling spectroscopy measurements on bilayer graphene on SiO$_2$ at 4.6 K. Our measurements show that with a back gate it is possible to tune the band gap and doping in the bilayer. The opening of the band gap is a promising effect for tailoring the bilayer for electronic applications. In addition, the shift of the Dirac point due to random impurities emphasizes the need to advance the sample preparation methods to ensure minimum impurities.

AD and BJL acknowledge the support of the U. S. Army Research Laboratory and the U. S. Army Research Office under contract/grant number W911NF-09-1-0333.  WB, ZZ and CNL acknowledge support by NSF CAREER DMR/0748910, NSF/ECCS 0926056, ONR N00014-09-1-0724 and ONR/DMEA H94003-09-2-0901.

\bibliography{Deshpande}

\end{document}